\documentclass[preprint,showpacs ,aps]{revtex4-1}
 \usepackage{amssymb,latexsym}
\usepackage{amsmath} 
\usepackage{graphicx} \usepackage{epsfig}
\usepackage{amssymb} \usepackage{color} \usepackage{slashed} 
\newcommand{\ba}{\begin{align}} \newcommand{\ea}{\end{align}}
\newcommand{\beq}{\begin{eqnarray}}
\newcommand{\eeq}{\end{eqnarray}} \newcommand{\nneeq}{\nonumber
\end{eqnarray}} \newcommand{\bs}{\boldsymbol}
 \newcommand{\nn}{\nonumber \\}
\newcommand{\es}{&=&}  \newcommand{\ps}{&+&}  
\newcommand{\ts}{& \times &} 
   
\newcommand{\bfig}{\begin{figure}}
\newcommand{\efig}{\end{figure}}

\newcommand{\cL}{ {\cal L} } 
 
\newcommand{\Fmn}{F_{\mu\,\nu}}

\newcommand{\gmu}{\gamma^\mu} 
\newcommand{\s}{ \slashed } 
 
 \newcommand{\ep}{\epsilon}

\begin{document} 


\title{\bf\large{Lightfront QED, Stueckelberg field and Infrared divergence}} 

\author{\bf T. R. Govindarajan $^{a,*}$, Jai D. More$^{b,\dagger}$ and P. Ramadevi$^{b,\,\dagger\dagger}$}  
\affiliation{$^{a}$ Chennai Mathematical Institute, Kelambakkam Siruseri, Tamil Nadu 600113, India. \\
 $^{b}$  Department of Physics,
  Indian Institute of Technology Bombay,\\ Powai, Mumbai 400076, India.}
  \email{trg@cmi.ac.in, trg@imsc.res.in, $^\dagger$ more.physics@gmail.com, $^{\dagger\dagger}$ramadevi@phy.iitb.ac.in }
\begin{abstract} 
Stueckelberg mechanism introduces a scalar field, known as Stueckelberg field, so that gauge symmetry is preserved in the massive abelian gauge theory. 
In this work, we show that the role of  the Stueckelberg field is similar to  the Kulish and Faddeev coherent state approach to handle infrared (IR) divergences. 
We expect that  the light-front quantum electrodynamics (LFQED) with Stueckelberg field must be IR finite in the massless limit of the  gauge boson. We have explicitly 
shown the cancellation of IR divergences in the relevant diagrams contributing to self-energy and vertex correction at  leading order.
\end{abstract} 

\maketitle

\section{Introduction}
It has been known for years, computation of
the transition matrix element in gauge theories like QED or QCD
inherits infrared divergences (IR) due to massless gauge boson
\cite{Bloch37, Yennie61, Naka58}. There are different remedies to cure IR
divergences. For example, one can treat these IR divergences by
mass regularization where a small mass is introduced for the gauge
particle. The other standard way is to perform dimensional
regularization. Or traditionally one introduces a small energy cutoff for the in
and out states, since the instruments come with natural
limitations. We find the divergence gets canceled as we remove
the cutoff \cite{Leibbrandt75}.
	
The standard Lehmann-Symanzik-Zimmermann (LSZ) formalism is based
on the assumption that at large time `$t > T$' the coupling
can be ``switched off" or in other words the particles can be treated as
free particles in the scattering process in the limit $| t | \rightarrow \infty$. Thus, in that case, the
initial and the final state can be considered as Fock state to
calculate the matrix element. However, in gauge theories with
massless particles, in particular, $U(1)$ gauge theory, the
asymptotic states are not the free states
. The charged particles
are dressed by soft or long wavelength photons as pointed out by
Kulish and Faddeev (KF) \cite{Kulish70}. Experimentally, these low
energy photons cannot be detected by the detector and lead to soft
divergences or IR divergences if one tries to do a theoretical
computation. The cancellation of IR divergences at amplitude level was first summarized by Chung \cite{Chung65}, which says that IR divergences can be eliminated if one chooses the initial and final
states to be charged particles with a suitable superposition
of an infinite number of photons.
KF demonstrated that the asymptotic interaction Hamiltonian in QED is
non-vanishing. They obtained the appropriate initial and final
state thereby modifying the Hilbert space, which is nothing but
the new asymptotic states. They constructed the asymptotic state
for QED by defining modified gauge invariant $S$-matrix and showed
that the IR divergences cancel at amplitude level using this new
basis.

Bagan {\it et al.} \cite{Bagan:2001wj} analogous to KF described
that the coupling in QED does not asymptotically vanish hence the
matter field is unphysical. Thus if one does not take into account
carefully the asymptotic behavior then the IR divergences appear.
They constructed a gauge invariant fermion wavefunction which handles the soft photon to obtain the IR finite result.

The KF method was later applied to obtain a set of asymptotic
states in the asymptotic region of perturbative quantum
chromodynamics (pQCD) by Nelson and Butler \cite{Nelson78}. It was
shown that the asymptotic states constructed leads to cancellation
of IR divergences in certain matrix elements in the lowest order in
pQCD. KF approach was utilized by various authors \cite{Greco78,
Dahmen81} to develop the coherent states in QED and QCD, in order
to study the IR behavior of abelian as well as non-abelian gauge
theories. The matrix elements using these coherent states were
shown to be IR finite.

The canonical field theory methods reviewed so far can also be analyzed in
light front (LF) formalism devised by Dirac \cite{Dirac49}. One of the advantages
of LF quantization is that it provides the understanding of Feynman's infinite momentum frame,
in which all finite mass particles behave like massless particles. The smooth
massless limit can be perceived in this context.
 	
Harindranath and Vary \cite{Hari88} applied coherent state
formalism to light front field theory (LFFT) for the first time
and showed that a coherent state may be a valid vacuum in LFFT. 
 A coherent state formalism was developed to deal with the true IR divergences in light front \cite{Misra94} and  later applied by one of us to deal with true IR 
divergences in light-front QED (LFQED) to calculate fermion mass
renormalization \cite{Jai12, Jai13, Jai15}. 
In the coherent state approach, the initial
and final Fock state is replaced by superposition of an infinitely
large number of soft photons in terms of new Hilbert space. It was shown
that IR divergences cancel up to $O(e^4)$ in LFQED using
coherent state basis in light front gauge \cite{Jai12} and Feynman gauge \cite{Jai13}. 
Later, this method was generalized for cancellation of IR divergence in
fermion mass renormalization to all orders in LFQED \cite{Jai15}.

It is known that IR divergences in gauge invariant QED 
is due to massless vector bosons. 
It will be interesting to look for alternate theories, which are free from IR divergences, giving physics of QED.
For example, consider the case of Stueckelberg formalism wherein an additional scalar field led to massive gauge boson with a
salient feature of gauge invariance \cite{Jac00, Ruegg04}. This
was in contrast to massive vector Proca field which was just the
extension of $U(1)$ gauge theory with the additional mass term with a
drawback of violation of abelian gauge symmetry. The additional
triumph of Stucekelberg theory was that it is a renormalizable theory
\cite{Ruegg04}. Not only the Stueckelberg mechanism, but there are other ways of generating vector boson mass like the well-known
Higgs mechanism in field theories and $B\wedge F$ terms as in  BF topological theories .

  Recently the infrared question and soft photon theorems have been
linked to new asymptotic symmetries that emerge for massless
particles in gauge theories \cite{Kapec17, Laddha18}. The
Stueckelberg QED has an additional degree of freedom which exists for 
the massive gauge bosons. Preserving the degrees of freedom at null 
infinity, while taking the limit of gauge boson mass to zero, we can 
get additional global or asymptotic symmetry. In fact, one of 
us \cite{Govindarajan19} discussed modified soft photon theorems due 
to massive photons and analyzed the subtle  procedure of 
taking massless limit.

 The paper is organized as follows: In Sec. \ref{SLag} we give the 
QED Lagrangian after the addition of Stueckelberg field and then 
obtain a generalized Stueckelberg Hamiltonian. In Sec. \ref{SE}, it is shown that IR divergences cancel when one takes 
the limit $m \rightarrow 0$ for the scalar field using light-front 
formalism up to leading order for self-energy correction. We 
also checked that IR divergences up to $O(g^3$) 
cancel in the massless limit of the scalar field $B$ is discussed in Sec. \ref{VC}. We conclude with our remarks
and future plans in Sec. \ref{conclusion}.
\section{Stueckelberg Lagrangian}\label{SLag}
We start by writing QED 
Lagrangian with Stueckelberg field \cite{Jac00,Ruegg04}
\beq
\cL\es \cL_{\psi}+\cL_{Stueck}+\cL_{g f}, 
\eeq 
where
\beq \cL_{\psi}\es\bar{\psi} \left[ \left(i\, \partial_\mu + g A_\mu
\right)\gmu-M\right] \psi, \\ \label{LStueck} 
\cL_{Stueck}\es -\frac1{4} \Fmn F^{\mu\,\nu}+\frac{1}{2}m^2\left(A_\mu -\frac{1}{m}\partial_\mu\, B \right)\left(A^\mu
-\frac{1}{m}\partial^\mu\, B \right)\\ \label{Lgf}
\cL_{g f}\es-\frac1{2\alpha}\left(\partial_\mu\, A^\mu+\alpha m
B\right)\left(\partial_\nu\, A^\nu+\alpha m B\right)
\eeq 
$\psi$, $A$ and $B$ describe the fermion field, gauge vector field and Stucekelberg scalar fields respectively. The field-strength tensor is given by: $F_{\mu \, \nu}=\partial_\mu A_\nu-\partial_\nu A_\mu$
The term in Eq. \ref{Lgf} is the gauge fixing term to remove the redundancy.
For simplification, we choose Feynman gauge which corresponds to
$\alpha = 1$.
It should be noted that the Stueckelberg field decouples to conventional QED in the mass going to zero limit. In fact, to start with,
electromagnetic potential $A_\mu$ has four components. The field
equations led to the massless particle, the photon with two
transverse physical degrees of freedom due to gauge invariance.
However, the addition of mass term spoils the gauge invariance. 
But, by introducing an extra scalar field $B$ we have five fields now.
This is Stucekelberg trick which gives Lorentz covariant and gauge
invariant massive spin-1 theory. 

The Lagrangian without the gauge fixing term is invariant under
the gauge transformation: 
\beq 
A_\mu \rightarrow A_\mu~+~\partial_\mu \lambda, ~~B~\rightarrow
B~+~m\lambda,~~\psi~\rightarrow e^{ig\, \lambda}\psi.
\eeq 
The complex gauge function satisfies the field equation congruent with $A_\mu$ and $B$;
\beq
(\partial^2+m^2) \lambda \es 0.
\eeq
The Stueckelberg field is not coupled to the fermion, which is
actually not a gauge invariant remark. We make the following gauge
transformation using the Stueckelberg field itself as gauge
parameter: 
\beq 
A_\mu \es \tilde{A}_\mu + \frac{1}{m}\partial_\mu\, B .
\eeq 
After this transformation we have the fermion field coupled to the
Stueckelberg field through derivative interaction. The Lagrangian
in the new variables is: 
\beq
 \cL\es \bar{\psi} \left[\gmu
\left(i\, \partial_\mu + g \tilde{A}_\mu +\frac{g}{m}\partial_\mu
B\right)-M\right] \psi - \frac1{4}
\tilde{F}_{\mu\nu}\tilde{F}^{\mu\nu}+\frac{1}{2}m^2\left(\tilde{A}_\mu \tilde{A}^\mu
\right)^2-\frac1{2}\left(\partial_\mu\, \tilde{A}^\mu\right)^2\nn \ps
\frac1{2}(\partial_\mu B)(\partial^\mu B)-\frac1{2}m^2\,B^2
\eeq
The fermion wavefunction $\psi$ can be decomposed into independent and dependent component of $\xi$ and $\eta$ respectively (for details {\it cf.} Ref \cite{Mus91}). 
\beq
 \psi= \xi+\eta
\eeq
The components of four vector $A_\mu$ are chosen as
$$ \tilde{A}_+=a_++\alpha_+, \qquad \tilde{A}_-=a_-=0,\qquad \tilde{A}_k=a_k
$$

This Hamiltonian can be obtained generalizing \cite{Jai12} for
massive photon QED in light front as
\beq 
P^-=H_0+V_1+V_2+V_3+V_B 
\eeq where

\begin{equation} 
H_0=\int
d^2 \bs{x}_\perp dx^-\bigg[\frac{i}{2}\bar \xi\gamma^-\partial_-\xi+\frac{1}{2}(F_{12})^2-\frac{1}{2}a_+\partial
_-\partial_ k a_k\bigg] \label{H_0} 
\end{equation}
 is the free Hamiltonian,
\begin{equation} V_1=g\int d^2 \bs{x}_\perp dx^-\bar
\xi\gamma^\mu\xi a_\mu
 \end{equation}
 is the $O(g)$, standard
3$-$point interaction vertex,
\begin{align}\label{V2} 
V_2=&g\int
d^2 \bs{x}_\perp dx^-\bar \eta\gamma^- \bar \partial_-\eta \nonumber\\
=&-\frac{i}{4}g^2\int d^2 \bs{x}_\perp dx^-dy^- \epsilon (x^--y^-)(\bar \xi a_k \gamma^k)(x)\gamma^+(a_j\gamma^j\xi)(y)
\end{align} 
is an $O(g^2)$ non-local effective 4$-$point vertex corresponding to
instantaneous fermion exchange and 
\begin{align}\label{V3}
V_3=&\frac{g}{2}\int d^2 \bs{x}_\perp dx^- \bar \xi
\gamma^+\xi\varphi_+\nonumber\\ 
=&-\frac{g^2}{4}\int d^2 \bs{x}_\perp
dx^-dy^-(\bar\xi \gamma^+\xi)(x) \vert x^--y^-\vert (\bar
\xi\gamma^+\xi)(y)
\end{align} 
is an $O(g^2)$ non-local effective
4$-$point vertex corresponding to an instantaneous photon
exchange.
\beq
 V_B\es \frac{g}{m}\int d^2\bs{x}_\perp\,
dx^-\bar{\xi}\gamma^\mu\xi\, \partial_\mu B 
\eeq
is the $O(g)$ 3$-$point interaction term due to the Stueckelberg field after gauge transformation.

$\xi$ and $a_\mu$ have standard expansions in terms of creation
and annihilation operators: 
\beq
\xi(x)\es \int\frac{d^2
{\bs p}_\perp}{(2\pi)^{3/2}}\int
\frac{dp^+}{\sqrt{2p^+}}\sum_{s=\pm\frac{1}{2}}[u(p,s)e^{-i(p^+x^--{\bs p}_\perp {\bs x}_\perp)} b(p,s,x^+) \nonumber\\
\ps v(p,s)e^{i(p^+x^--{\bs p}_\perp {\bs x}_\perp)}d^\dagger(p,s,x^+)], \\
 a_\mu(x)\es\int\frac{d^2{\bs k}_\perp}{(2\pi)^{3/2}}\int
 \frac{dk^+}{\sqrt{2k^+}}\sum_{\lambda=1,2}\epsilon^\lambda_\mu(kq)[e^{-i(k^+x^--{\bs k}_\perp {\bs x}_\perp)}a(k,\lambda,x^+)\nonumber\\
\ps e^{i(k^+x^--{\bs k}_\perp {\bs x}_\perp)}a^\dagger(k,\lambda,x^+)],\\
\partial_\mu B(x)\es \int\frac{d^2{\bs q}_\perp}{(2\pi)^{3/2}}\int \frac{dq^+}{\sqrt{2q^+}}\sum_{\lambda'=1,2}q_\mu
[e^{-i(q^+x^--{\bs q}_\perp {\bs x}_\perp)}a(q,\lambda',x^+)\nonumber\\
\ps e^{i(q^+x^--{\bs q}_\perp {\bs x}_\perp)}a^\dagger(q,\lambda',x^+)]
\eeq
 and satisfy 
\begin{align}\label{anticommutation}
\{b(p,s),b^\dagger(p^\prime,s^\prime)\}&=\delta(p^+-p^{\prime+})\delta^2({\bs p}_\perp-{\bs p}^\prime_\perp)\delta_{ss^\prime}\nonumber\\
&=\{d(p,s),d^\dagger(p^\prime,s^\prime)\},\nonumber\\
[a(k,\lambda),a^\dagger(k^\prime,\lambda^\prime)]&=\delta(k^+-k^{\prime+})\delta^2({\bs k}_\perp-{\bs k}^\prime_\perp) \delta_{\lambda\lambda^\prime}.
\end{align}

These commutation/anticommutation relations hold at equal
light-front time $x^+$.

\section{Self energy correction up to $O(g^2)$}\label{SE}
 We obtain the
transition matrix element in light-front time ordered perturbation
theory using perturbative expansion as follows \beq T\es V + V
\frac{1}{p^--H_0}V+ \cdots \eeq \begin{figure}[h]
\includegraphics[scale=0.4]{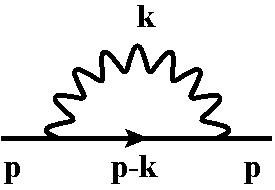} \hspace{2cm}
\includegraphics[scale=0.4]{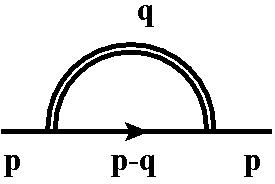} 
\caption{Self energy diagram. In the figure on the left, wavy lines indicate massive gauge boson while the double line (in the figure on right) represents the Stueckelberg field. The fermion field is denoted by a straight line.}
 \label{selfenergy}
  \end{figure}

The contribution to self-energy $O(g^2)$ correction is obtained
from
 \beq
T^{(1)}(p,\,p) \es
T_{1a}(p,\,p)+T_{1b}(p,\,p)+T_{1c}(p,\,p)\nn \label{VV}
\es \langle p,s \vert
V_1 \,\frac{1}{p^--H_0}\,V_1 \vert p,s \rangle +\langle p,s \vert
V_B\,\frac{1}{p^--H_0}\,V_B \vert p,s \rangle +\langle p,s \vert
V_2 \vert p,s\rangle 
\eeq
 In Eq. \ref{VV} on the right
hand side, we obtain the $O(g^2)$ contributions to fermion
self-energy correction. The first term corresponds to the standard
three-point vertices, the second term corresponds to three-point
vertices due to the Stueckelberg field while the third term corresponds
to four-point instantaneous vertex which arises in light front
quantization. We will focus on the true IR divergences in the
massless limit which shows up due to the vanishing energy
denominator. The third term can be dropped as it is not
contaminated by IR divergence. It is very important to understand
here that in the second term the energy denominator gets IR
divergence, when one takes $m \rightarrow 0$ limit. Then the
second term contribute when longitudinal polarization is used,
equal and opposite to the first term. In order to calculate the
transition matrix element $T_{1a}$ and $T_{1b}$ contributing to
fermion self-energy correction to $O(g^2)$, we insert a
complete sets of states to account for the intermediate state.
Details of the calculation follow: \beq T_{1a}(p,\,p)\es
\sum_{spins}\int \prod_{i=1}^2 d^3p_i^\prime d^3k_i^\prime \langle
p,s \vert V_1 \vert
p_1^\prime,s_1^\prime,k_1^\prime,\lambda_1^\prime\rangle\langle
p_1^\prime,s_1^\prime,k_1^\prime,\lambda_1^\prime\vert\frac{1}
{p^- - H_0}\vert
p_2^\prime,s_2^\prime,k_2^\prime,\lambda_2^\prime\rangle \nn \ts
\langle p_2^\prime,s_2^\prime, k_2^\prime,\lambda_2^\prime\vert
V_1\vert p,s \rangle \eeq On substituting for $V_1$ and further
simplification gives 
\beq T_{1a}(p,\,p) \es \frac{g^2}{2(2
\pi)^3\, p^+}\int{d^2{\bs k}_{1\perp}}\int\frac{dk^+}{k^+}
\frac{(p\cdot \epsilon(k))^2}{(p^--p_1^--k^-)}\nn \es
-\frac{g^2}{2(2 \pi)^3\, p^+}\int{d^2{\bs
k}_{\perp}}\int\frac{dk^+}{k^+p_1^+}
\frac{Tr[\epsilon\llap/^\lambda(k,\,\lambda) (\not p_1+m)
\epsilon\llap /^\lambda(k,\,\lambda) (\not
p+m)]}{4(p^--p_1^--k^-)}\label{trace1}\\ 
\es-\frac{g^2}{2(2 \pi)^3\, p^+}\int{d^2{\bs
k}_{\perp}}\int\frac{dk^+}{k_1^+p_1^+} \frac{Tr[\s k (\s
p_1+m) \s k (\s p+m)]}{4\,M^2\,(p^--p_1^--k^-)}\label{trace2}
\eeq In going from Eq. \ref{trace1} to Eq. \ref{trace2} we have
written the longitudinal polarization vector as \cite{Banks08}
\beq
\ep_\mu(k,\,\lambda)\es \frac{k_\mu}{m}+O(\frac{\mu}{k})+\cdots
\label{longitudinal} \eeq Since we focus on IR divergence in the
massless limit which come from the disappearance of the
longitudinal mode we use corresponding polarization.

In the similar manner, we can calculate the diagram in Fig.
\ref{selfenergy}(b) \beq T_{1b}(p,\,p)\es\langle p \vert \, V_B\,
\frac{1} {p^- - H_0}\, V_B \,\vert p\rangle \nn 
\es\frac{g^2}{2(2 \pi)^3\, p^+}\int{d^2{\bs q}_{\perp}}\int\frac{dq^+}{q^+p_1^+} \frac{Tr[\s q (\s p_1+m) \s q (\s p+m)]}{4\,M^2\,(p^--p_1^--q^-)}\label{trace3}
\eeq 
In the limit $m \rightarrow 0$, $k=q$ and we observe that
IR divergences in Eq. \ref{trace2} and Eq. \ref{trace3} cancel
exactly each other.

It was shown that IR divergences cancel when we use coherent state
basis instead of Fock state to calculate the same matrix element in Ref \cite{Jai12}. 
Now, we have calculated self-energy correction up to $O(g^2)$ and
shown that the IR divergences cancel when we use Stueckelberg field
and take $m \rightarrow 0$ limit. This explains up to $O(g^2)$ the
contribution for the terms responsible for the cancellation for IR divergences in the coherent state basis are provided by the
Stueckelberg field.
 
\section{Vertex correction up to $O(g^3)$}\label{VC}
In this section, we discuss the lowest order radiative correction for 3-point interaction
in light front formalism. It was shown in Ref \cite{Misra94} the IR divergences cancel when one uses
coherent state basis instead of Fock state to calculate the matrix
element. Now we compute using the Stueckelberg field.

The $O(g^3)$ correction terms due to the three-point vertex contributing to IR divergences are given by
\begin{figure}[h] 
\includegraphics[scale=0.3]{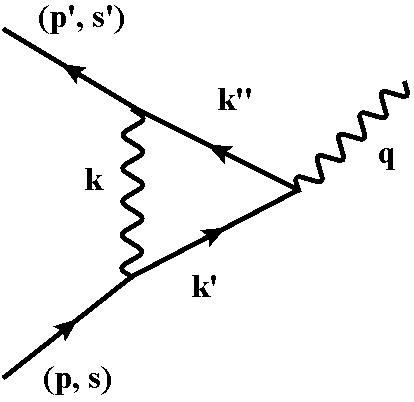}
\hspace{1cm} \includegraphics[scale=0.3]{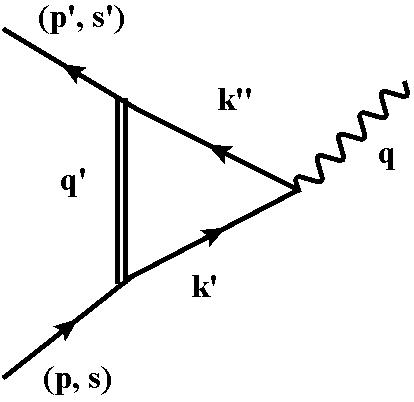} 
\caption{The relevant vertex correction diagrams contributing to IR divergence up to $O(g^3).$ } 
\label{vertex}
\end{figure} 
\beq
T^{(1)}_{12}(p',p,q) \es
T_{12\,(a)}+T_{12\,(b)}+T_{12\,(c)}+T_{12\,(d)}\\
 \es \langle p',s',\, q,\lambda \vert \, V_1\, \frac{1} {p^- - H_0}\, V_1 \,
\frac{1} {p^- - H_0}\, V_1 \,\vert p,s\rangle \nn
\ps \langle p',s',\,
q,\lambda \vert \, V_B\, \frac{1} {p^- - H_0}\, V_B\, \frac{1}
{p^- - H_0}\, V_1 \,\vert p,s\rangle
\nn
 \ps \langle p',s',\, q,\lambda \vert \, V_2\, \frac{1} {p^- - H_0}\, V_1 \,\vert
p,s\rangle + \langle p',s',\, q,\lambda \vert \, V_3\, \frac{1}
{p^- - H_0}\, V_1 \,\vert p,s\rangle 
\eeq 
where
$T_{12(a)}$ corresponds to vertex correction contribution coming from three 3-point vertices $V_1$. 
As we have an additional three-point vertex $V_B$ due to Stueckelberg field interaction with the fermion field which is represented by $T_{12(b)}$.
The subscript $12$ corresponds to one particle state going to two particle states. 
There are also contributions for vertex correction coming from the vertex due to instantaneous fermion vertex $V_2$ and instantaneous boson vertex $V_3$ corresponding to $T_{12(c)}$ and $T_{12(d)}$  respectively. We consider only the diagrams shown in Fig. \ref{vertex} and also we limit our calculation for the vertex correction  $\Lambda^+(p',\,p)$ hence the last term do not
contribute due to the tensor structure.

The contribution to vertex correction up to $O(g^3)$ is given by
\beq 
T_{12\,(a)}\es\ep_\mu \Lambda_{21(a)}^\mu \nn \es
g^3\int\frac{[dk]}{k^+\,k_1^+\,k_2^+}\frac{\bar{u}(p',s')\,\gamma^\alpha\,(\s
k_1+m)\,\gamma^\mu\,(\s
k_2+m)\,\gamma^\beta\,u(p,\,s)\ep_\alpha(k,\lambda)\,\ep_\beta(k,\lambda)\ep_\mu(q, \lambda')}{(p^--k^--k_1^-)(p^--k^--k_2^--k'^-)}\nn
\es \frac{g^3}{M^2}\int\frac{[dk]}{k^+\,k_1^+\,k_2^+}\frac{Tr\left[\s k \,(\s k_1+m)\,\s k'\,(\s k_2+m)\,\s k\right]}{(p^--k^--k_1^-)(p^--k^--k_2^--k'^-)} \\
T_{12\,(b)}\es\ep_\mu \Lambda_{21(b)}^\mu \nn 
\es-\frac{g^3}{M^2}\int\frac{[dq']}{q'^+\,k_1^{'+}\,k_2^{'+}}\frac{\bar{u}(p',s')\,\s
q'\,(\s k_1'+m)\,\gamma^\mu\,\,(\s k_2'+m)\,\s
q'\,u(p,\,s)\ep_\mu(q, \lambda')}{(p^--q'^--k_1^{'-})(p^--q'^--k_2^{'-}-k'^-)}
\nn 
\es-\frac{g^3}{M^2}\int\frac{[dq']}{q'^+\,k_1^{'+}\,k_2^{'+}}\frac{Tr\left[\s
q'\,(\s k_1'+m)\,\s k'\,(\s k_2'+m)\,\s
q'\right]}{(p^--q'^--k_1^{'-})(p^--q'^--k_2^{'-}-k'^-)} 
\eeq

We use again longitudinal polarization from Eq. \ref{longitudinal}. We observe the addition of the Stueckelberg field leads to the
cancellation of IR divergences in the limit $m \rightarrow 0$ at
the amplitude level itself as in the self-energy computation.
Coherent states led to the similar cancellation of IR divergence
was shown earlier in Ref \cite{Misra94}. Again it is clear the role of
soft photon in coherent states is played by Stueckelberg field. Using
coherent states IR divergence cancellation was extended to all
orders in Ref \cite{Jai15}. We do not anticipate
difficulty in establishing similar cancellation using Stueckelberg
field.

\section{Conclusion} \label{conclusion}
In this work, we have shown that IR divergences get cancelled 
if one adds Stueckelberg field to QED lagrangian using light-front 
formalism. Massive Stueckelberg QED has been studied earlier 
and shown to be renormalizable. For very low mass it has also been
shown to reproduce results of conventional QED as long as the 
interaction is through  a conserved current. 
Our main goal was to study  $m~\rightarrow~0$ limit in Stueckelberg QED where IR divergences could make its appearance. 
Interestingly, we could reproduce the leading order results expected 
by KF approach. It appears that the arguments can be extended to establish the cancellation of IR divergences to all orders.
We hope to pursue in future  such a generalization where the tools discussed in Ref. \cite{Jai15} will be useful.

Applying the Stueckelberg mechanism to QCD has limitations due to self 
coupling amongst gauge bosons. In fact, the non-abelian Stueckelberg 
theory is  non-renormalisable. Hence, the problem of achieving IR 
divergence cancellation in QCD needs a novel approach. The recent 
paper \cite{Glazek2018} attempts IR issues through Higgs mechanism 
in abelian theory. Comparing their methods with our approach may 
give us a tangible idea to handle IR problems in non-abelian theories.   

Another interesting aspect to explore will be the interplay 
between IR divergences and supersymmetry in 
the context of supersymmetric formulation of Stueckelberg QED  
\cite{Kors05}. Also it is known that SUSY Stueckelberg QED is 
renormalizable.

We have confined to gauge theories in this work. It will be 
challenging to investigate Stueckelberg mechanism in gravity theories.
For example, the linearised  massless gravity theory studied by
van Dam, Veltman, and Zakharov (vDVZ) has 
a discontinuity known as vDVZ discontinuity\cite{Hinterbichler12}. 
The addition of new fields into the massive gravity theory could 
remove such discontinuities. These new fields, similar to the 
Stueckelberg field, are
required so that the number of degrees of freedom are unchanged
even after taking the massless limit. These 
additional Stueckelberg degrees of freedom can play non-trivial
role due to its gravitational interaction. This will be presented 
elsewhere {\cite{TRG}.

\begin{acknowledgements}
 The authors would like to acknowledge Prof V. Ravindran for 
extremely useful discussions. 
J.D.M would like to thank Prof V. Ravindran and T.R.G for warm 
hospitality during her visit to IMSc. J.D.M wishes to thank P.R 
for financial support through the IRCC funded project by 
IIT Bombay no. 17 RPA001.
\end{acknowledgements}

\end{document}